# Further Discussions on Sufficient Conditions for Exact Relaxation of Complementarity Constraints for Storage-Concerned Economic Dispatch

Zhengshuo Li, Qinglai Guo, Hongbin Sun, Jianhui Wang

*Abstract*—Storage-concerned economic dispatch (ED) problems with complementarity constraints are strongly non-convex and hard to solve because traditional Karush-Kuhn-Tucker (KKT) conditions do not hold in this condition. In our recent paper, we proposed a new exact relaxation method which directly removes the complementarity constraints from the model to make it convex and easier to solve. This paper further extends our previous study, with more than one group of sufficient conditions that guarantee the exact relaxation presented, proven and discussed. This paper may contribute to wider application of the exact relaxation in storage-concerned ED problems.

## I. INTRODUCTION

THE use of energy storage systems is being widely considered in power generation economic dispatch (ED) [1]. However, complementarity constraints, which prevent simultaneous charging and discharging of energy storage devices, should be included in a storage-concerned ED model, making the model strongly non-convex and hard to solve efficiently. Mathematically, with the complementarity constraints considered, the storage-concerned ED problem as a kind of the so-called "mathematical programs with equilibrium constraints" (MPEC) is greatly different from conventional non-convex problems, because the traditional Karush-Kuhn-Tucker (KKT) conditions are invalid and conventional methods, e.g., interior-point method, cannot be directly applied [2]. Regarding this issue, mixed-integer programming (MIP) methods are usually used in practical engineering fields [3]-[4], with exact penalty methods or smoothing methods [5]-[8], and regularization relaxation methods [9] also investigated. In MIP methods, binary variables are introduced to transform the complementarity constraints into (linear) mixed-integer one; while in the other three methods, a series of conventional non-convex problems should be iteratively solved to approach the "true" optimal solution of the MPEC. However, these methods may all result in long solution times due to additional integer variables or excessive iterations. Exact relaxation methods were recently proposed in [10] and [11], based on an additional assumption that charging storage does not affect or decrease the total operational cost of the grid. Although the methods neither introduce additional variables nor lead to iterations, they may be challenged under conditions where storage owners should pay the grid for charging energy, or the energy constraints of storage are active. It is thus necessary to investigate if the complementarity constraint can be "exactly" relaxed for general cases. (Here "exact" means that, despite relaxation, the same optimal solution will always be obtained.) In our recent paper [12], a new exact relaxation method is proposed, where the complementarity constraints are directly removed from the model to make it convex so as to be solved much more efficiently. The validity of the new exact relaxation method is also mathematically proven in [12].

Compared with [12], this paper further extends the previous study, with two more new groups of sufficient conditions proposed and proven to guarantee the exact relaxation, and the differences from the one in [12] discussed as well. This paper may contribute to wider application of the exact relaxation method in storage-concerned ED problems.

## II. STORAGE-CONCERNED ED MODEL

In order to make better understanding of the exact relaxation method and the new sufficient conditions for exact relaxation, the storage-concerned ED model in [12] is cited and listed below.

Consider a power system that has $N$ buses and $L$ lines, and the dispatch horizon lasts from $t = 1$ to $t = T$, with a time interval $\Delta t$. Taking operational cost functions of the generators and storage as inputs, a direct current (DC) model based storage-concerned ED model [13] can be formulated below,

$$\min F = \sum_{i \in N} \sum_{t \in T} \left( g_i \left( P_i^{dc}(t) \right) - f_i \left( P_i^{ch}(t) \right) \right) + \sum_{i \in N} \sum_{t \in T} h_i \left( P_i^G(t) \right) \quad (1)$$

subject to the following constraints for any $t \in T$, $i \in N$, and $j \in L$:

$$0 \leq P_i^{ch}(t) \leq \bar{P}_i^{ch}(t) , \qquad \alpha_{i,1}(t), \alpha_{i,2}(t) \quad (2)$$

Zhengshuo Li, Qinglai Guo and Hongbin Sun are all with the Department of Electrical Engineering, State Key Laboratory of Power Systems, Tsinghua University, Beijing 100084, China (e-mail:shb@tsinghua.edu.cn).
Jianhui Wang is with Argonne National Laboratory, Argonne, IL, USA.

$$0 \leq P_i^{dc}(t) \leq \bar{P}_i^{dc}(t), \qquad \alpha_{i,3}(t), \alpha_{i,4}(t) \quad (3)$$

$$E_i(t) = (1-\varepsilon_i)^t E_i^0 + \sum_{\tau=1}^{t} (1-\varepsilon_i)^{t-\tau} \left(\eta_i^{ch} P_i^{ch}(\tau) - P_i^{dc}(\tau)/\eta_i^{dc}\right) \Delta t \quad (4)$$

$$E_i^{\min}(t) \leq E_i(t) \leq E_i^{\max}(t), \qquad \beta_{i,1}(t), \beta_{i,2}(t) \quad (5)$$

$$\sum_{t=1}^{T} (1-\varepsilon_i)^{T-t} \left(\eta_i^{ch} P_i^{ch}(t) - P_i^{dc}(t)/\eta_i^{dc}\right) \Delta t \geq E_i^r, \quad \varphi_i \quad (6)$$

$$P_i^{ch}(t) P_i^{dc}(t) = 0 \quad (7)$$

$$\underline{P}_i^G \leq P_i^G(t) \leq \bar{P}_i^G \quad (8)$$

$$R_i^{dn} \Delta t \leq P_i^G(t+1) - P_i^G(t) \leq R_i^{up} \Delta t \quad (9)$$

$$\sum_{i \in N} P_i^G(t) + \sum_{i \in N} \left(P_i^{dc}(t) - P_i^{ch}(t)\right) = \sum_{i \in N} D_i(t), \quad \lambda(t) \quad (10)$$

$$\underline{P}_j^{Ln} \leq \sum_{i \in N} GSF_{j-i} \left(P_i^G(t) + P_i^{dc}(t) - P_i^{ch}(t) - D_i(t)\right) \leq \bar{P}_j^{Ln},$$

$$\mu_{j,1}(t), \mu_{j,2}(t) \quad (11)$$

where $g_i$ is the discharging cost of the storage device at bus $i$; $f_i$ is the storage charging fee. $f_i > 0$ means storage pays grid for charging, and vice versa if $f_i < 0$; $h_i$ is the operational costs of the generator at bus $i$; $P_i^{ch}(t)$ and $P_i^{dc}(t)$ are the grid-side charging and discharging power of the storage at time $t$; $\bar{P}_i^{ch}(t)$ and $\bar{P}_i^{dc}(t)$ are the rated limits of the charging and discharging power; $\eta_i^{ch}$ and $\eta_i^{dc}$ are charging and discharging efficiencies; $E_i(t)$ is the stored energy at time $t$; $E_i^{\min}(t)$ and $E_i^{\max}(t)$ are the lower and upper limits of the stored energy; $E_i^0$ is the initial energy; $\varepsilon_i$ is the self-discharge rate; $E_i^r$ is the total charging demand of the storage device; $P_i^G(t)$ is the output of the generator at time $t$; $\underline{P}_i^G$ and $\bar{P}_i^G$ are the lower and upper limits of the output; $R_i^{up}$ and $R_i^{dn}$ are the ramp limits of the generator. $D_i(t)$ is the load at bus $i$ at time $t$; $GSF_{j-i}$ is the generation shift factor to line $j$ from bus $i$. $\bar{P}_j^{Ln}$ and $\underline{P}_j^{Ln}$ are the upper and lower limits of the transmission capacity of the line; $\lambda(t)$ is the multiplier of the equalities, and $\alpha_{i,1}(t)$ to $\alpha_{i,4}(t), \beta_{i,1}(t), \beta_{i,2}(t),$ and $\mu_{j,1}(t), \mu_{j,2}(t), \varphi_i$ are the non-negative multipliers of the corresponding inequalities.

The model can be described as follows. The objective in (1) is the total operational cost of the generators and the storage. Usually, $h_i$ is modeled as a convex quadratic function, $g_i$ is a convex non-decreasing function, and $f_i$ is a linear function, so that the objective is convex. As a special case, if the storage is owned by the power system, $g_i$ and $f_i$ can be zero. Constraints (2) and (3) are the rated charging and discharging power limits of storage; Constraint (4) is the integral relationship between the stored energy and the prior charging and discharging process from $\tau = 1$ to $t$, with self-discharge and a round-trip efficiency considered as [10]; In (5), the energy limit is modeled, which equivalently represents the state-of-charge limit of the storage; Constraint (6) represents the net charging requirements of the storage device – especially when it is an aggregation of electric vehicles (EVs), $E_i^r$ is the total charging demand. Notably, if the charging requirement need not be considered in the model, then the value of $E_i^r$ can be taken as a very large negative number (e.g., -100000), making the constraint (6) actually relaxed in the optimal solution. Constraint (7) is the complementarity constraint, which makes the problem strongly non-convex and KKT conditions invalid. Moreover, if the storage is an aggregation of EVs, the upper and lower limits of (2), (3), (5) can be time-varying, so timestamps are used there. Constraints (8) and (9) describe the generation and ramp limits of a generator; Constraint (10) is the power balance of the power system at time $t$, and (11) describes the bidirectional power flow limits of transmission lines.

Notably, although the storage-related cost term in the objective function is modeled as $\sum_{i \in N} \sum_{t \in T} \left(g_i(P_i^{dc}(t)) - f_i(P_i^{ch}(t))\right)$, it doesn't mean that the model is only limited to the conditions where storage pays the grid for charging. Actually, the objective function and the dispatch model represent at least the following three storage dispatch scenarios with positive or negative discharging prices $g_i'$ and charging prices $f_i'$, which are listed and explained in Table I.

TABLE I
THREE SCENARIOS WHERE THE OBJECTIVE FUNCTION AND THE MODEL CAN BE APPLIED

| Scenario | Description | Signs of ($f_i'$, $g_i'$) |
|---|---|---|
| Scenario 1 | Both charging and discharging are costs for grid's dispatch | ($f_i' < 0$, $g_i' > 0$) |
| Scenario 2 | Storage operational cost is neglected in grid's dispatch | ($f_i' = 0$, $g_i' = 0$) |
| Scenario 3 | Storage pays the grid for charging energy and the grid pays the storage for discharging energy | ($f_i' > 0$, $g_i' > 0$) |

From Table I, it can be seen that since the sign of $f_i$ is assumed "−" in the objective function, positive charging price ($f_i' > 0$) means that the charging payment is income to the grid; while negative charging price ($f_i' < 0$) means that the charging payment is cost to the grid. Following that, the three scenarios in Table I are explained below:

- In Scenario 1, it is assumed that both charging and discharging are regarded as costs to grid's dispatch, which is also the scenario considered in [10]. As explained above, this scenario can be represented by setting $f_i' < 0$, in the objective function in (1).
- Scenario 2, as a special case where storage charging and discharging costs are not considered in ED, is considered in [11]. This scenario might happen when the storage device is owned by the power grid company itself.
- Scenario 3 is a very common scenario considered in literature [14]-[21], especially when the storage is an aggregation of electric vehicles, e.g., in [14]-[17]. In this scenario, the storage pays the grid for charging energy and get money from the grid for discharging. From the viewpoint of the grid, the dispatch problem in this scenario can be regarded as one kind of maximizing social welfare problems, as shown in [22] (but we wrote it in an equivalent minimization problem form). Notably, if the complementarity constraint (7) is relaxed in this scenario, then charging and discharging may well happen simultaneously, which will be further discussed in the following part.

In all, it can be seen that the established storage-concerned ED model in [12] is a general model which can be applied to many typical scenarios considered in literature. Hence, an exact relaxation method based on this general model has potential to be applied in general cases.

## III. RELAXATION CONDITIONS AND PROOF

If constraint (7) is relaxed, i.e., removed from the model, then the relaxed model (RM) becomes convex so that the global optimal solution can be easily obtained (note that any local optimal solution is globally optimal for convex problems, and KKT conditions are both necessary and sufficient). Since KKT conditions are valid for RM, let $L$ denote the Lagrangian function of the RM, $\xi_i$ denote $1-\varepsilon_i$, and $\Gamma(t) = \sum_{\tau \geq t} \xi_i^{\tau-t} \left( \beta_{i,1}(\tau) - \beta_{i,2}(\tau) \right) + \xi_i^{T-t} \varphi_i$. Then, using KKT conditions, the following equations

$$\frac{\partial L}{\partial P_i^{ch}(t)} = -f_i'\left(P_i^{ch}(t)\right) - \alpha_{i,1}(t) + \alpha_{i,2}(t) - \eta_i^{ch} \Gamma(t) \Delta t \\ + \lambda(t) + \sum_j GSF_{j-i} \left( \mu_{j,1}(t) - \mu_{j,2}(t) \right) = 0 \quad (12)$$

and

$$\frac{\partial L}{\partial P_i^{dc}(t)} = g_i'\left(P_i^{dc}(t)\right) - \alpha_{i,3}(t) + \alpha_{i,4}(t) + \Gamma(t) \Delta t / \eta_i^{dc} \\ - \lambda(t) - \sum_j GSF_{j-i} \left( \mu_{j,1}(t) - \mu_{j,2}(t) \right) = 0 \quad (13)$$

hold.

With (12) and (13), it can be proven that the relaxation is exact under the following groups of sufficient conditions.

### A. Sufficient Conditions of Group A (Conditions A)

Conditions A have been shown and proven in [12], as follows: for any $i \in N$,

***Condition A-1***:

$$g_i'\left(P_i^{dc}(t)\right) \geq f_i'\left(P_i^{ch}(t)\right), \forall t \quad (14)$$

***Condition A-2***:

$$f_i'\left(P_i^{ch}(t)\right) < LMP_i(t), \forall t \quad (15)$$

where $LMP_i(t) = \lambda(t) + \sum_j GSF_{j-i} \left( \mu_{j,1}(t) - \mu_{j,2}(t) \right)$ denotes the location marginal price (LMP) at bus $i$ at time $t$ in a DC model.

The proof of the exact relaxation under Conditions A can be proven by contradiction, and the details can be referred to [12].

The physical meaning of Condition A-1 is that the marginal compensation paid to them for discharging a unit of energy must cover the costs for the owner to charge that amount of energy back. Condition A-2 means that the storage charging price should be strictly less than the LMP at the connected location. Note that, Conditions A can also be applied to the cases with $f_i' \leq 0$ in [10] and [11]. Further detailed interpretation of Conditions A can be referred to [12]. [1]

Moreover, Conditions A are usually easy to check in field operation given the charging and discharging prices as inputs and the LMPs predicted with acceptable accuracy.

### B. Sufficient Conditions of Group B (Conditions B)

Conditions B are as follows: for any $i \in N$,

***Condition B-1***:

$$g_i'\left(P_i^{dc}(t)\right) > f_i'\left(P_i^{ch}(t)\right), \forall t \quad (16)$$

***Condition B-2***:

$$f_i'\left(P_i^{ch}(t)\right) \leq LMP_i(t), \forall t \quad (17)$$

The proof of Conditions B shown below is similar to that of Conditions A.

***Proof***

Assume that there exists $P_i^{ch}(t) > 0$ and $P_i^{dc}(t) > 0$ for storage $i$ at time $t$ in the optimal solution of the RM. Then $\alpha_{i,1}(t) = 0$, $\alpha_{i,3}(t) = 0$ because of the complementary slackness conditions. With $\alpha_{i,1}(t) = 0$, $\alpha_{i,2}(t) \geq 0$, and Condition B-2, it follows from (12) that $\Gamma(t) \geq 0$ holds for storage $i$ at time $t$.

Then, with $\alpha_{i,3}(t) = 0$, by combining (12) and (13),

$$\left(1/\eta_i^{dc} - \eta_i^{ch}\right)\Gamma(t)\Delta t + g_i' - f_i' + \alpha_{i,2}(t) + \alpha_{i,4}(t) = 0 \quad (18)$$

holds. Given $\alpha_{i,2}(t)$, $\alpha_{i,4}(t) \geq 0$, $1/\eta_i^{dc} - \eta_i^{ch} > 0$ and Condition B-1, it can be inferred that $\Gamma(t) < 0$ also holds for storage $i$ at time $t$, so there is a contradiction.

Therefore, it can be seen that $P_i^{ch}(t) > 0$ and $P_i^{dc}(t) > 0$ cannot both appear in the optimal solution of the RM for any storage at any time slot. Hence, the relaxation is exact under sufficient Conditions B. □

Note that different from Conditions A, the charging price in Conditions B can be equal to the LMP and the exact relaxation still holds if Condition B-1 is satisfied. Although that difference may be not so mathematically striking, it extends the potential application of the exact relaxation method. For example, if storage charging is regarded as regular loads and charged at the LMP, which violates Condition A-2, then Conditions A are invalid but Conditions B can be used to determine the exact relaxation as long as the discharging compensation price is strictly larger than charging price.

### C. Sufficient Conditions of Group C (Conditions C)

Conditions C are as follows: for any $i \in N$,

***Condition C-1***:

$$g_i'\left(P_i^{dc}(t)\right) > f_i'\left(P_i^{ch}(t)\right)/\eta_i^{cycle}, \forall t \quad (19)$$

***Condition C-2***:

$$LMP_i(t) \geq 0, \forall t \quad (20)$$

where $\eta_i^{cycle} = \eta_i^{ch}\eta_i^{dc} < 1$ denotes the round-trip efficiency of the storage.

The proof is also completed by contradiction.

***Proof***

Assume that there exists $P_i^{ch}(t) > 0$ and $P_i^{dc}(t) > 0$ for storage $i$ at time $t$ in the optimal solution of the RM. Then $\alpha_{i,1}(t) = 0$, $\alpha_{i,3}(t) = 0$ because of the complementary slackness conditions.

By $\eta_i^{cycle} \times (13) + (12)$, it follows that

---

[1] At the end of the third last paragraph of Section III of [12], the inequality "inf $g_i'(P_i^{dc}(t)) \leq \sup f_i'(P_i^{ch}(t))$" was incorrectly printed. "≤" should have been "≥".

$$\left(\eta_i^{cycle} g_i'\left(P_i^{dc}(t)\right) - f_i'\left(P_i^{ch}(t)\right)\right) \\ + \left(1-\eta_i^{cycle}\right) LMP_i(t) + \alpha_{i,2}(t) + \eta_i^{cycle}\alpha_{i,4}(t) = 0 \quad (21)$$

Given Conditions C, it follows that

$$\alpha_{i,2}(t) + \eta_i^{cycle}\alpha_{i,4}(t) < 0. \quad (22)$$

As $\alpha_{i,2}(t), \alpha_{i,4}(t) \geq 0$, there is a contradiction, so the relaxation is exact. □

The physical meaning of Condition C-1 is that the discharging price paid to the storage owners must cover the charging price weighted by the reciprocal of round-trip efficiency. Condition C-2 means that the shadow price of the storage-connected location should be non-negative, which was usually true in past years but is being challenged in current grids with more and more renewable energies integrated. When Condition C-2 holds, the increase of loading tends to increase the whole operational costs of the grid, so the energy loss brought by simultaneous charging and discharging pattern becomes uneconomic and suboptimal in terms of ED. Hence, there must be no simultaneous charging and discharging happening in the optimal solution of the relaxed ED model.

## IV. DISCUSSIONS ABOUT THE SUFFICIENT CONDITIONS

### A. Impact of LMP prediction on the conditions

LMPs used in the above sufficient conditions are more difficult to forecast in comparison with traditional price forecasting due to the transmission congestion on top of other factors (weather, load, etc.) In spite of the difficulty, researchers have kept studying LMP forecasting and reported effective approaches [23]-[27], e.g., neural networks based approaches, fuzzy inference system based approaches, and a load probability distribution based approach. In [28], it is reported that with an artificial neural network (ANN) approach, the mean absolute percent error (MAPE) of the LMP is from 0.9% to 1.5% with different load patterns. For better illustration of the prediction accuracy, some results from [28] are cited and shown in Fig. 1.

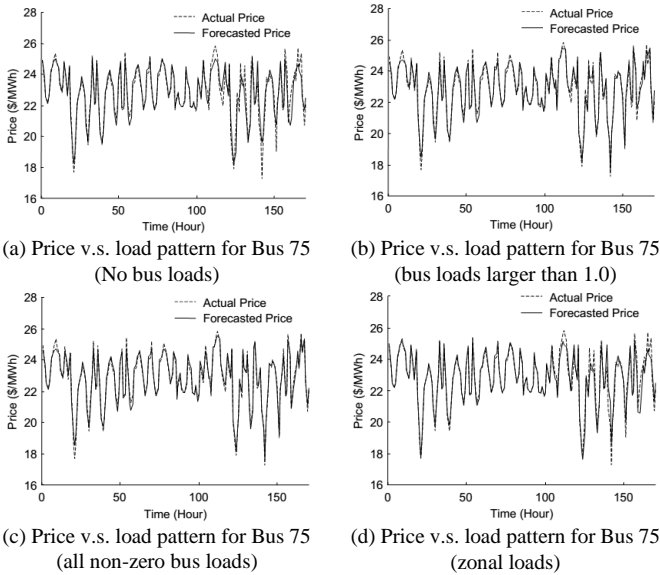

(a) Price v.s. load pattern for Bus 75 (No bus loads)
(b) Price v.s. load pattern for Bus 75 (bus loads larger than 1.0)
(c) Price v.s. load pattern for Bus 75 (all non-zero bus loads)
(d) Price v.s. load pattern for Bus 75 (zonal loads)

Fig. 1 Citation of the LMP prediction results with the ANN approach in [28]

Moreover, besides using the forecasted LMPs, the lower bound of the interval where the actual LMP may fall, denoted by $\underline{LMP}_i(t)$, can also be used in the above sufficient conditions to determine the exactness of the relaxation in field operation. Specifically, if charging price $f_i'$ is below the lower bound $\underline{LMP}_i(t)$, then it can be inferred that Condition A-2 would be very likely satisfied. Similarly, if the lower bound $\underline{LMP}_i(t)$ is non-negative, then Condition C-2 would be very likely satisfied.

In fact, using $\underline{LMP}_i(t)$ instead of $LMP_i(t)$ helps to widen the application of the exact relaxation, because it is usually easier to predict the interval where LMPs may probably fall than the accurate value of the LMPs in real world. Estimation of $\underline{LMP}_i$ can be as follows: with MAPE as an approximate estimation of the standard deviation of LMP forecasting, we have the lower bound

$$\underline{LMP} = (1 - 3 \times MAPE) \times \text{forecasted LMP} \quad (23)$$

Taking Fig. 1 as an example, with the forecasted LMP is 20 $/MWh and MAPE = 1%, the lower bound of the LMP is estimated as 19.4, and then Condition C-2 must be satisfied very likely. Moreover, if the charging price is much less than the lower bound (e.g., charging price is 10 $/MWh), it can be inferred that Condition A-2 must be satisfied very likely as well.

Notably, given the uncertainty in LMPs, the condition where Condition B-2 is binding (namely, charging price = LMP), though academically interesting, becomes a little too ideal for practical application, for it is rather hard to predict the exact value of the LMP and determine if the inequality in (17) active.

### B. Necessity of the LMP-related sufficient conditions

Note that every group of the sufficient conditions has a sufficient condition related to LMPs. Although the LMP-related condition might appear trivial in form, it is really necessary to guarantee the exactness of the relaxation.

Taking Scenario 3 in Table I into consideration, without LMP-related sufficient conditions, simultaneous charging and discharging can happen in the optimal solution of the RM. For example, suppose a certain storage charges and discharges simultaneously in the optimal solution of the RM, and the overall dispatch effect is that the storage is delivering power to grid, which can be described by $P_i^{dc}(t) > 0$, $P_i^{ch}(t) > 0$, and $P_i^{dc}(t) - P_i^{ch}(t) > 0$. Consider a new solution with $\hat{P}_i^{dc}(t) > 0$ and $\hat{P}_i^{ch}(t) = 0$ satisfying $\eta_i^{ch} P_i^{ch}(t) - P_i^{dc}(t)/\eta_i^{dc} = \eta_i^{ch}\hat{P}_i^{ch}(t) - \hat{P}_i^{dc}(t)/\eta_i^{dc}$ (so that it is also feasible for storage charging/ discharging process), then it follows that $\hat{P}_i^{dc}(t) = P_i^{dc}(t) - \eta_i^{cycle} P_i^{ch}(t)$. Hence, the overall dispatch effect with regard to $\hat{P}_i^{dc}(t)$ and $\hat{P}_i^{ch}(t)$ is $\hat{P}_i^{dc}(t) - \hat{P}_i^{ch}(t) = P_i^{dc}(t) - \eta_i^{cycle} P_i^{ch}(t)$ which is larger than $P_i^{dc}(t) - P_i^{ch}(t)$, so the generators' output is decreased with the new solution, and so is the generators' operational cost. Nevertheless, since the storage charging payment to the grid is also decreased by $f_i\left(P_i^{ch}(t)\right)$ at the same time, it cannot be directly determined whether the new solution is better than the original one with $P_i^{dc}(t) > 0$, $P_i^{ch}(t) > 0$ so as to avoid the simultaneous charging and discharging happen in the optimal solution of the RM. In other words, the LMP-related condition, either balancing the marginal charging income and LMP, or evaluating the signature of LMP, helps to guarantee for a decreased objective value with $\hat{P}_i^{dc}(t)$ and $\hat{P}_i^{ch}(t)$ so that simultaneous charging and discharging cannot happen in the optimal solution of the RM. Therefore, the LMP-related condition importantly contributes to the exact relaxation, especially for Scenario 3. Related numerical results can be referred to [12], where a case with $g_i' = 25 > f_i' = 24$ satisfies Condition A-1 but violates Condition A-2. It can be found that the maximum of $P_i^{ch}(t)P_i^{dc}(t)$ of the RM is over 0.12 in that case, indicating the relaxation is not exact. Therefore, the LMP-related condition in each group of conditions is necessary to be considered and checked.

*C. Application range comparison*

Firstly, as stated above, if LMPs can be accurately predicted, Conditions B has a wider application range than Conditions A. However, given the uncertainty in LMPs, the application ranges of Conditions A and B are almost the same.

Then, comparison between Conditions A and C is discussed. Obviously, Condition C-1 appears more stringent than Condition A-1, but it can be expected that both conditions may well be satisfied in reality, because almost no storage owners would participate in the power grid dispatch otherwise. Hence, the difference of Conditions A and C mainly lies at the difference of the LMP-related conditions.

As stated above, LMPs are usually positive in traditional power grids. In that condition, Conditions C have a wider application range than Conditions A, because there is no need to check if charging price is less than LMPs, and the exactness of the relaxation can be directly determined only with Condition C-1. Moreover, given Condition C-1 almost always satisfied, it can be concluded that the exactness always holds in that condition.

However, with wind generation integrated, LMPs may be negative (For instance, as reported in [21], the western zone of the ERCOT sometimes faced negative LMPs), so Condition C-2 is violated and Conditions C become invalid. However, if $f_i'\left(P_i^{ch}(t)\right) < 0$, which means storage charging is also rewarded by the grid[2], then Condition A-2 is still satisfied as long as $f_i'\left(P_i^{ch}(t)\right) < LMP_i < 0$, and then exactness of the relaxation is determined using Conditions A. Hence, it can be seen that Conditions A have a wider application range than Conditions C in this condition.

In a word, if the prediction LMPs are positive, then Conditions C are recommended for use; otherwise, Conditions A or B are recommended for use in filed operation.

---

[2] This may happen when storage charging helps to absorb more spilled wind and maintain the energy balance of the grid.

## V. SIMULATIONS

Most simulation platform information has been described in [12]. Major simulation results and discussions can be also found there. Only the parameters and simulation results not shown in [12] due to the space limitations are listed here.

The test system in [12] is an IEEE 30-bus system provided by MATPOWER [29], where 5 units are connected @ buses 1, 13, 22, 23, 27, and a wind farm is connected @ bus 2. The parameters of the units are shown in Table II. Note that the constant term in $h_i(x)$ has no impacts on the results, so it is not considered. The 50 storages are connected at the PQ buses (each bus has two storage facilities on average), and the charging/discharging rates and storage capacity can be referred to [12]. The self-discharging rate is 0.01/hour [10]. The total loads and the available wind power are shown in Fig. 2.

TABLE II
PARAMETER OF THE UNITS

| Unit No. | $\underline{P_i^G}$ (MW) | $\overline{P}_i^G$ (MW) | $R_i^{up}$ (MW/15min) | $R_i^{dn}$ (MW/15min) | $h_i(x)$ |
|---|---|---|---|---|---|
| 1 | 20 | 100 | 5 | -5 | $0.04x^2 + 10x$ |
| 2 | 40 | 100 | 5 | -5 | $0.01x^2 + 20x$ |
| 3 | 20 | 80 | 8 | -8 | $0.02x^2 + 23x$ |
| 4 | 20 | 120 | 6 | -6 | $0.01x^2 + 22x$ |
| 5 | 20 | 120 | 6 | -6 | $0.01x^2 + 10x$ |

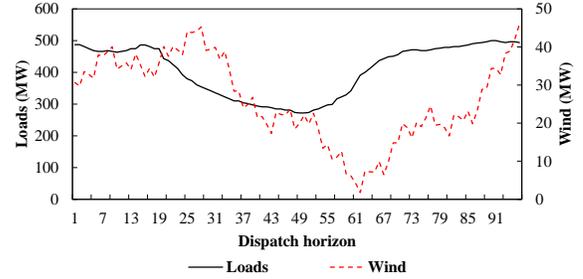

Fig. 2  The total loads and maximum wind output during the dispatch horizon

Furthermore, the optimal dispatch of the overall storage power is shown in Fig. 3.

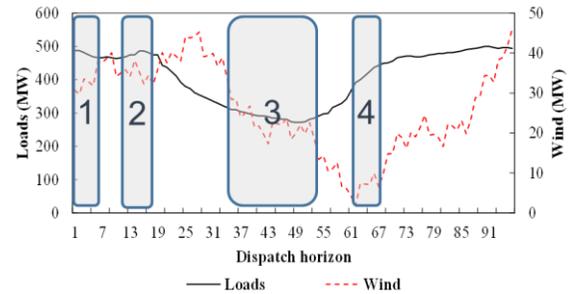

(a) Load curve in the dispatch horizon

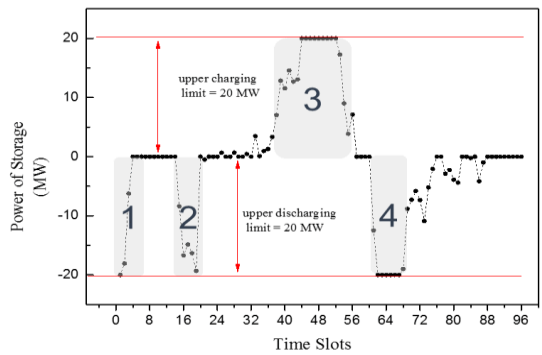

(b) Optimal dispatch of the overall storage power in the dispatch horizon

Fig. 3  Optimal dispatch of the overall storage power

From the numbered grey areas in Fig. 3, it can be seen that storage charges mostly at the load valley (e.g., grey area 3) and discharges most at the peak load time (e.g., grey areas 1 and 2) or scarce wind time (e.g., grey area 4), which coincides with the anticipation of the system operators.

## VI. Conclusion

This paper which extends our previous work in [12] proposes and analyzes two more groups of sufficient conditions for exact relaxation of a storage-concerned ED problem. Besides one group of the sufficient conditions in [12], two more groups of sufficient conditions are also proven, discussed and compared here, with several other important issues which are not or not fully discussed in [12] further investigated. This paper not only helps with better understanding of the exact relaxation method proposed in [12], but also contributes to a wider range of the application of the exact relaxation method in future filed grid operation with storage and renewable energies widely integrated.